\documentclass{iaus}
\usepackage{graphicx}

\title[~~Low Metallicity ISM] 
{Low Metallicity ISM: excess submillimetre emission and CO-free H$_{2}$ gas}

\author[Madden et al ]   
{Suzanne C. Madden$^1$, Aur\'elie R\'emy $^1$, Fr\'ed\'eric Galliano $^1$, Maud Galametz$^2$, George Bendo$^3$, Diane Cormier$^1$, Vianney Lebouteiller$^1$,  Sacha Hony$^1$  and {\it Herschel} SAG 2 consortium }

\affiliation{$^1$CEA Saclay, DSM, AIM, Service d'Astrophysique, \\ Gif-sur-Yvette 91911,
France, email: {\tt suzanne.madden@cea.fr} \\ 
{$^2$Institute of Astronomy, University of Cambridge, Madingly Rd., Cambridge,
UK }\\
 {$^3$Alma Regional Center, University of Manchester , Oxford Rd., Machester, UK \\ [\affilskip]}
}

\newcommand{\hers}{{\it Herschel}}

\newcommand{\spit}{{\it  Spitzer}}
\newcommand{\iso}{{\it ISO}}
\newcommand{\iras}{{\it IRAS}}

\newcommand{\mic}{$\mu$m}

\newcommand{\mics}{\mu {\rm m}}

\newcommand{\bet}{$\beta$}

\pubyear{2011} 
\volume{284}  
\pagerange{1--12}
\jname{The Spectral Energy Distribution of Galaxies}
\editors{R.J. Tuffs \&  C.C.Popescu, eds.}
\begin{document}

\maketitle

\begin{abstract}
 
The low metallicity interstellar medium of dwarf galaxies gives a different picture in the far infrared(FIR)/submillimetre(submm)wavelengths than the more metal-rich galaxies. Excess emission is often found in the submm beginning at or beyond 500 \mic. Even without taking this excess emission into account as a possible dust component, higher dust-to-gas mass ratios (DGR) are often observed compared to that expected from their metallicity for moderately metal-poor galaxies.
 The SEDs of the lowest metallicity galaxies, however, 
give very low dust masses and excessively low values of DGR, inconsistent with the amount of metals expected to be captured into dust if we presume the usual linear relationship holding for all metallicities, including the more metal-rich galaxies. This transition seems to appear near metalllicities of 12 + log(O/H) ~ 8.0 - 8.2. These results rely on accurately quantifying the total molecular gas reservoir, which is uncertain in low metallicity galaxies due to the difficulty in detecting CO(1-0) emission. Dwarf galaxies show an exceptionally high [CII] 158 \mic/CO (1-0) ratio which may be indicative of a significant reservoir of 'CO-free' molecular gas residing in the photodissociated envelope, and not traced by the small CO cores. 

\keywords{galaxies: dwarf, galaxies: ISM, ISM: molecules}
 \end{abstract}

\firstsection 
\section{Introduction}
 
Understanding how galaxies evolve requires fundamental comprehension of the process of star formation and the subsequent effects on the interstellar medium (ISM) under conditions in the early universe - an epoch when the ISM had not yet endured many cycles of chemical enrichment.   We can attempt to extrapolate to such conditions, by studying the interplay between star formation and dust and gas in low metallicity dwarf galaxies, of which our local universe hosts a veritable zoo.  

The first systematic study of dust in dwarf galaxies observed with wavelengths as long as 100~\mic\ with \iras, showed lower infrared luminosity (L$_{IR}$) compared to H$\alpha$ than spirals. L$_{12 \mics}$/L$_{25\mics}$ values were lower than spirals, while the higher L$_{60\mics}$/L$_{100\mics}$ highlighted the presence of warmer dust and enhancement of very small grains (e.g. \cite[Hunter et al 1989]{hunter89}; \cite[Melisse \& Israel 1994]{melisse94}). The \iso\ and \spit\ missions opened up the window onto MIR spectroscopy, demonstrating the dearth of PAHs in dwarf galaxies, a consequence of the prevailing hard radiation field, shocks or delayed injection by AGB stars (e.g. \cite[Madden et al 2006]{madden06}; \cite[Wu et al 2006]{wu06}; \cite[O'Halloran et al 2008]{ohalloran08}; \cite[Galliano et al 2008]{galliano08}) as well as cooler dust revealed by longer wavelengths, extending to 160 \mic\ (e.g. \cite[Popescu et al 2002]{popescu02}). SCUBA/JCMT and Laboca/APEX observations exposed excess emission in dwarf galaxies at 850/870 \mic\ which, if interpreted as cold dust, could comprise a large amount of dust in low metallicity galaxies   compared to their measured gas reservoirs and their low metallicities (e.g. \cite[Galliano et al 2003]{galliano03}; \cite[Galliano et al 2005]{galliano2005}; \cite[Zhu et al 2009]{zhu09}; \cite[Galametz et al 2011]{galametz11}).

 One constraint on the dust mass should be the measured dust-to-gas mass ratio (DGR), if we understand how this parameter is controlled through chemical evolution. Atomic gas is often widespread in dwarf galaxies, in contrast to molecular gas, if the CO(1-0) is a reliable tracer of the molecular reservoir in these galaxies. On the other hand, the 158 \mic\ [CII], often assumed to arise from the photodissociation regions (PDRs) around molecular clouds, is excessively bright in dwarf galaxies compared to the CO, in contrast to more metal-rich galaxies (\cite[Poglitsch et al 1995]{poglitsch95}; \cite[Israel et al 1996]{israel96}; \cite[Madden et al 1997]{madden97};  \cite[Madden 2000]{madden99}). What is this telling us about the total molecular gas reservoir, the galaxy morphology and the distribution of the various gas phases in dwarf galaxies? 

\begin{figure} 
 \includegraphics[width=4.0 cm.,height=5.0cm.]{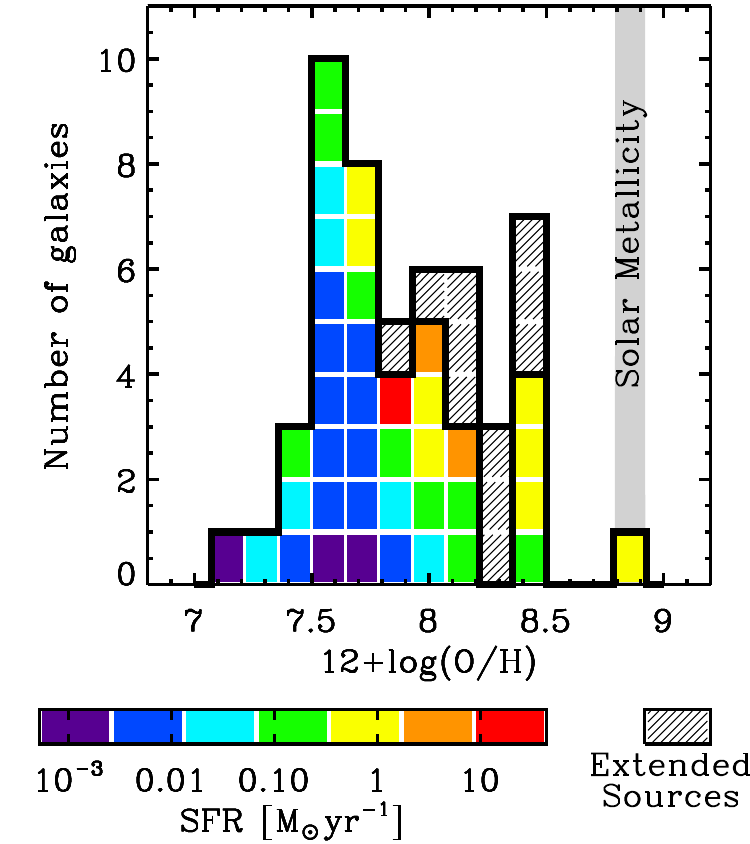} 
\hspace*{1.0 cm}
\includegraphics[width=7.0 cm.]{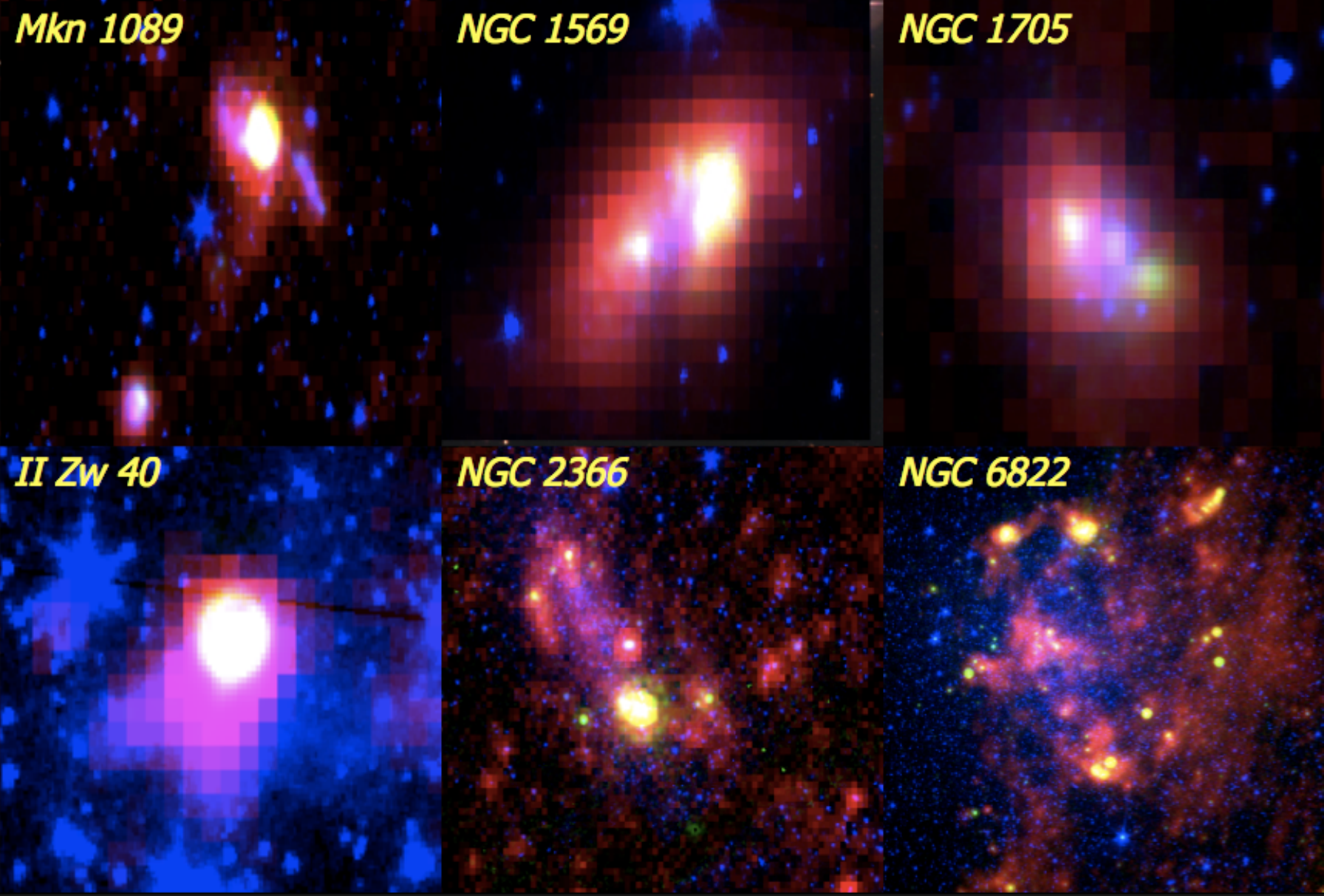} 
\caption{(left) Metallicity and star formation properties of the Herschel Dwarf Galaxy Survey. (right) \spit\ and \hers\ 3-color images of some well-studied survey dwarf galaxies. Blue: 3.6 \mic\ {\it IRAC/Spitzer} (stars); Green 24 \mic\ {\it MIPS/Spitzer} (hot dust); Red: 250 \mic\ {\it SPIRE/Herschel} (cool dust). Note the extended diffuse 250 \mic\ dust emitting throughout the galaxies, while the warmer 24 \mic\ dust is confined to the compact star forming sites. }
\label{fig1}
\end{figure}

 \hers\ has opened up the submm wavelength window beyond 160 \mic, with observations covering the 50 to 500 \mic\ window. 
The Dwarf Galaxy Survey (DGS) is a \hers\ key program targeting 48 local universe dwarf galaxies with a wide range of star formation properties and metallicity values, as low as 1/50 Z$\odot$ (Fig.\,\ref{fig1}) and as nearby as the Magellanic Clouds to study the multiphase components of the stars, gas and dust under low metallicity environments.
   
\section{SED models of dwarf galaxies}
PACS observations cover 3 photometric bands: 70, 100 and 160 \mic\ (FWHM $\sim$ 10"; \cite[Poglitsch et al 2010]{poglitsch10}) while SPIRE observes at 250, 350 and 500 \mic\ (FMHM=18" to 38"; \cite[Griffin et al 2010]{griffin10}). Comparison of the \spit\ 24 \mic\ images (\cite[Bendo et al 2012]{bendo12}) with the PACS 70 \mic\ and SPIRE 250 \mic\ (Fig.\,\ref{fig1}) highlights the extent of the cooler dust traced by the 250 \mic\ emission in contrast to the warmer dust emitting at 24 \mic, favoring the more compact HII regions. Coverage at \spit\ and \hers\ wavelengths provides well-sampled SEDs for wide-ranging studies of the dust properties in dwarf galaxies.
 
\begin{figure}
\begin{center}
 \vspace*{-1.3 cm}
\includegraphics[width=13cm.]{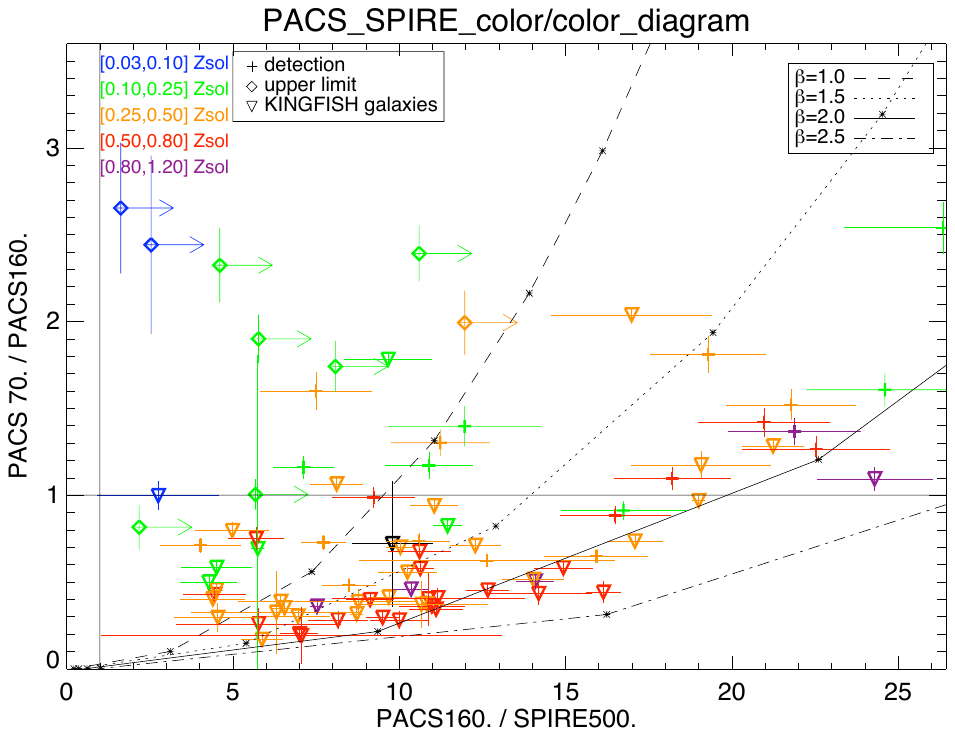} 
\vspace*{-7.9 cm}
 \caption{Herschel color-color diagram of DGS galaxies detected in at least all PACS bands: PACS70/160 and PACS160/SPIRE500. For comparison, KINGFISH galaxies (\cite[Dale et al 2012]{dale12}) are also shown (downward triangles).  Modeled modified black body fits with \bet\ = 1.0, 1.5, 2.0, 2.5 are shown in curves, with temperature values noted as dots on the curves, increasing from 10 K in steps of 10 K (as black dots) starting from the lower left of the curves to the upper right. Metallicity bins of the galaxies are shown in color.}
  \label{fig2}
\end{center}
\end{figure}

\begin{figure}
 \hspace*{0.0cm}
\includegraphics[width=5.1 cm.]{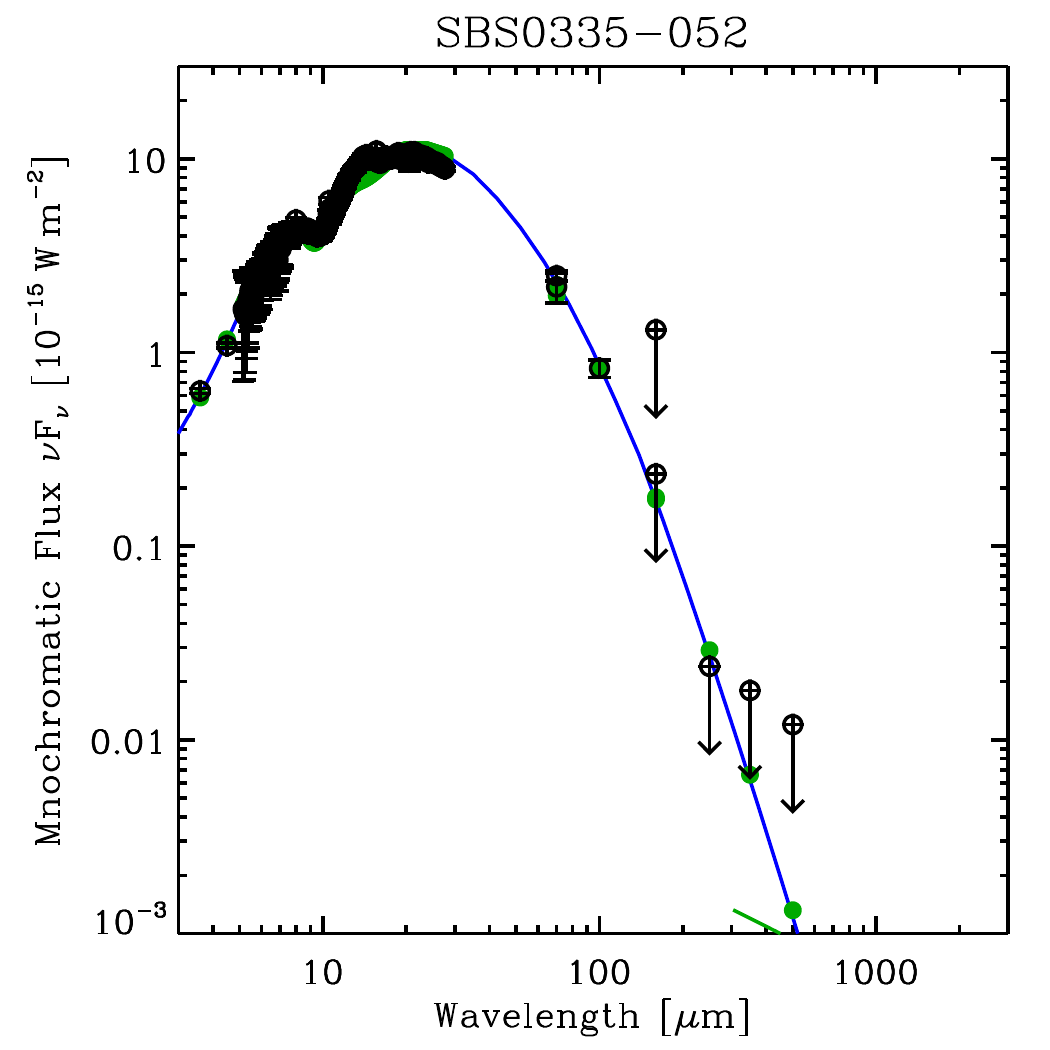} 
\hspace*{-1.0cm}
\includegraphics[width=5.1 cm.]{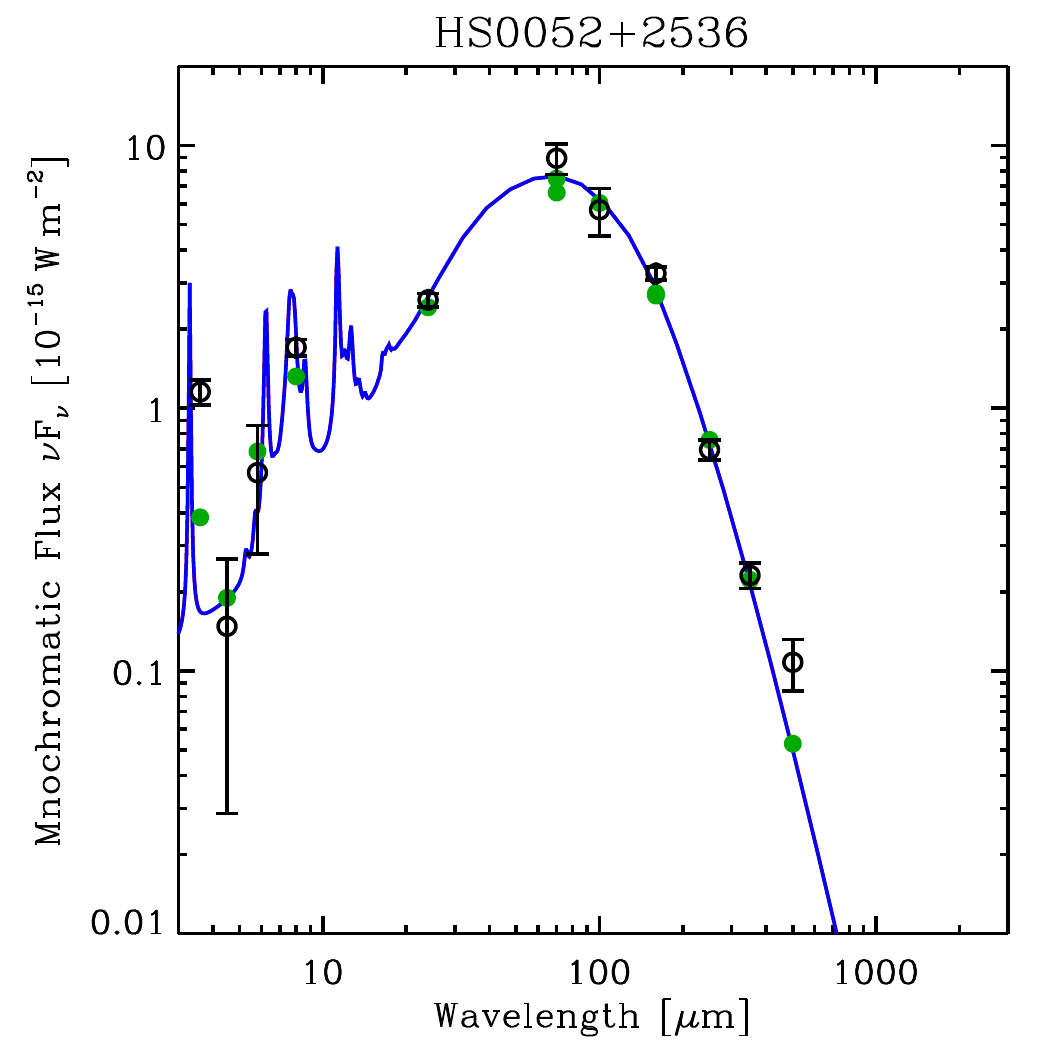} 
 \hspace*{1.2cm}
\includegraphics[width=5.1 cm.]{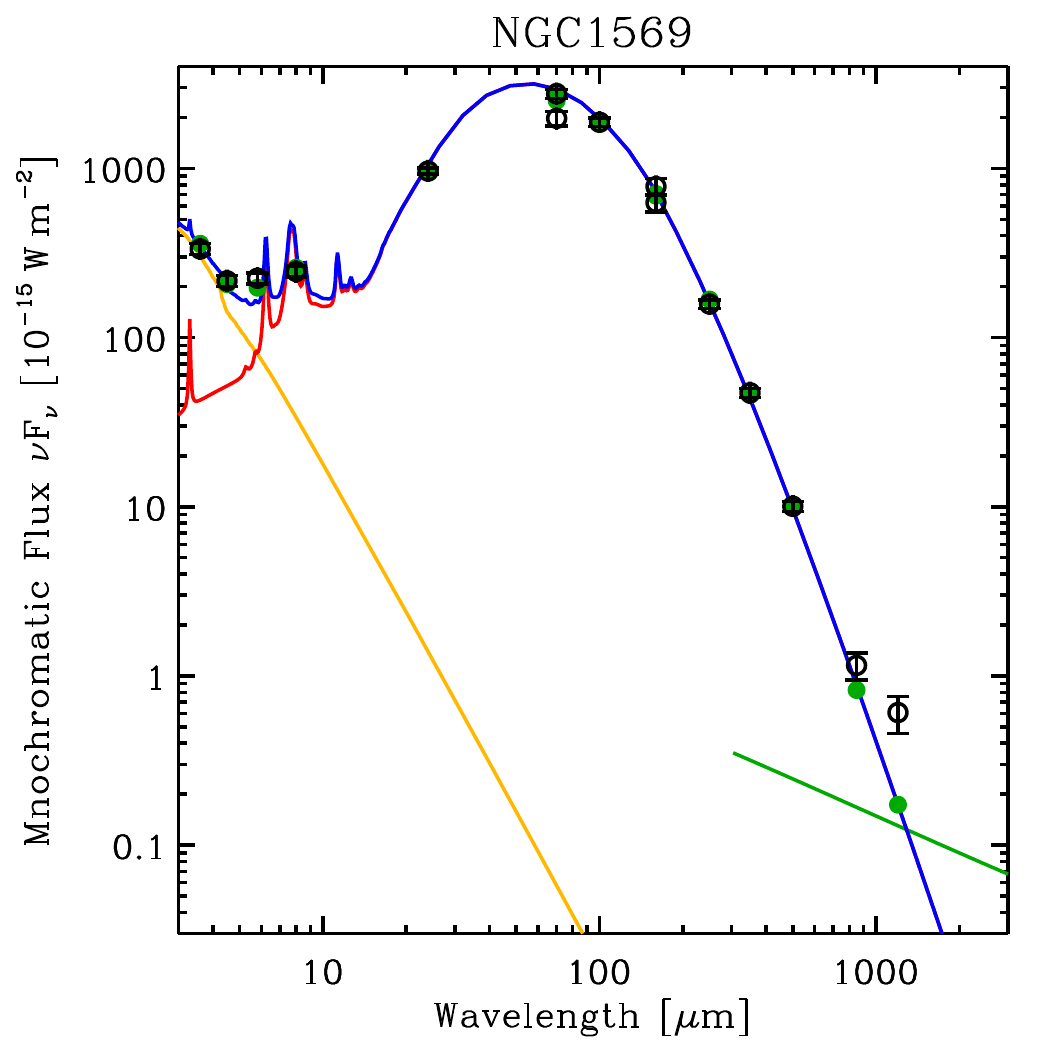} 
\hspace*{1.4cm}
\includegraphics[width=5.1 cm.]{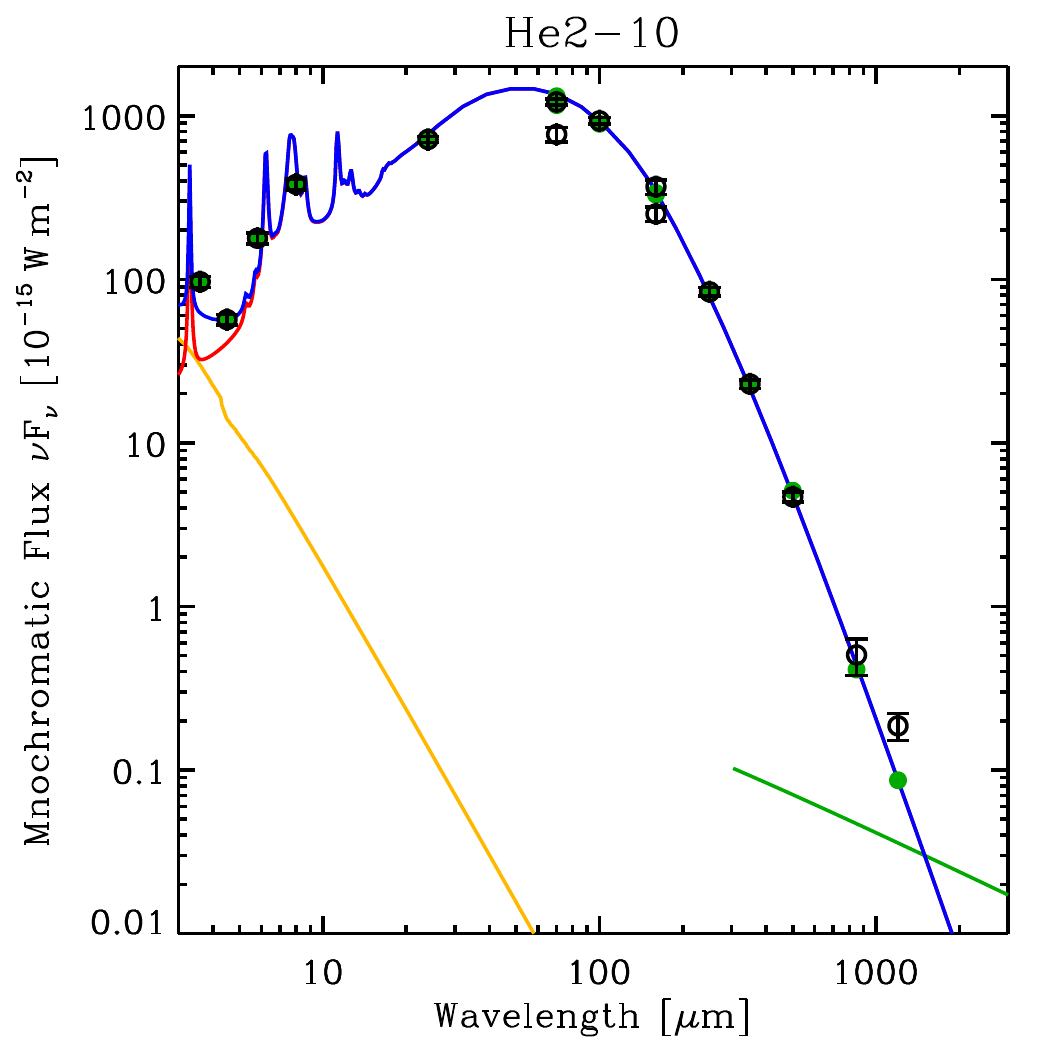} 
 
\caption{SEDs of dwarf galaxies. SBS0335-052 has on average very hot dust, with the SED peaking between 20 and 30 \mic. HS0052+2536 shows a very low \bet\ in the \hers\ color-color diagram (Fig.\,\ref{fig2}) and does indeed show a 500 \mic\ excess. NGC~1569 and He2-10 both fall near \bet\  $\sim$ 2 in the color-color diagram, but their full SED model unveils an excess beyond 500 \mic. \spit\ and \hers\ data are used to constrain the SED models. \hers\ observations are used for the model constraints in the cases of data redundancy (70 and 160 \mic). The yellow curve is the stellar contribution, the red curve the dust contribution, the green curve is the free-free contribution, extrapolated from the observed radio observations when available and the blue is the total expected SED. The green points show the predicted model fluxes.}
  \label{fig3}
 \end{figure}

{\underline{\it Metallicity and \bet\ and T effects}}. 
The FIR-submm behavior of the DGS sample is investigated via \spit\ and \hers\ color-color diagrams, to obtain an overview of the total sample (R\'emy et al in preparation; also this volume). Fig.\,\ref{fig2} shows the effect of varying emissivity indices (\bet) and average dust temperatures (T) presuming a single modified black-body to explain the FIR-submm emission. For comparison the KINGFISH galaxies (\cite[Dale et al 2012]{dale12}) are also included. The effect of the overall higher metallicity clusters the bulk of the KINGFISH sample between \bet~=1.5 and 2.0.
The most active dwarf galaxies, particularly the blue compact dwarfs, peak at wavelengths less than 70 \mic\ and some between 35 and 60 \mic, much shorter wavelengths compared to the more metal-rich starburst galaxies. The difference in the shape of the SEDs of dwarf galaxies can be noted via the \hers\ color-color diagram:  as a consequence of their overall 
hotter dust peaking at MIR wavelengths (higher PACS70/PACS160 values for the lower metallicity bins), their Rayleigh-Jeans slope drops off at FIR and submm wavelengths. Consequently, these galaxies are often not detected at all SPIRE wavelengths for the lowest metallicity galaxies, particularly at 500 \mic.  A prominent example of this effect is illustrated in one of the lowest metallicity galaxies of our sample, SBS0335-052E (12+1og(O/H) = 7.29; Fig.\,\ref{fig3}). The integrated SED shows the flatter MIR-FIR emission peaking between 20 and 30 \mic\ (\cite[Houck et al 2004]{houck04}; \cite[Galliano et al 2008]{galliano04}; Sauvage et al in preparation). The very dense super star clusters dominate the warm overall dust emission, with the observed SED leaving little evidence for cold submm-emitting dust.  

For those dwarf galaxies detected at both PACS and SPIRE wavelengths, modified black-body fits (omitting 70 \mic\ in the fits) give a wide range of \bet\ and T solutions with a mean \bet\ of 1.6 and mean T of 32 K (Fig.\,\ref{fig4}).  
Since the most metal-poor galaxies of the DGS sample do not have 500 \mic\ detections, this distribution of modeled \bet\ and T is primarily representing galaxies with metallicity values greater than 12 +log (O/H) $\sim$ 8.0 - 8.2.  For comparison, the KINGFISH galaxies show a similar mean \bet\ but the T distributions are cooler, peaking between $\sim$ 20 and 25 K (\cite[Dale et al 2012]{dale12}). Galaxies requiring an exceptionally low \bet\ solution, may be indicative of the presence of a submm excess which could start to be detectable at wavelengths as low as 500 \mic. 

{{\underline{\it  Submm excess examples}}. Taking into account the optical to submm and radio wavelengths, full SED models (\cite[Galliano et al 2008]{galliano08}) can help to further interpret the Herschel color-color diagrams.  For example, from the \hers\ color-color diagram (Fig.\,\ref{fig2}), HS0052+2536 has an exceptionally low \bet\ value $\sim 1$. Inspection of the fully modeled SEDs (Fig.\,\ref{fig3}) shows that a submm excess is indeed present.  About 50\% of the DGS galaxies detected at 500 \mic\ show a submm excess of $\sim$ 7\% to 100\% above the SED model (R\'emy et al, in preparation). 
 A relationship between metallicity and submm excess within the DGS sample is not yet obvious. This may be somewhat due to the requirement for 500 \mic\ detections which often omits the lowest metallicity galaxies. While this \hers\ color-color diagram highlights potential galaxies with submm excess, galaxies for which the excess begins beyond 500 \mic\ will be missed without observations at longer wavelengths. For example,  NGC~1569 and He2-10,  show \bet\ $\sim$ 2 in the Herschel color-color diagram but do indeed show a submm excess when including their ground-based 850 \mic\ observations (Fig.\,\ref{fig3}; see also \cite[Galliano et al 2003]{galliano03}; \cite[Galliano et al 2005]{galliano2005}; \cite[Galametz et al 2011]{galametz11}). 
 
{{\underline{\it DGRs and possible origins of the submm excess}}. Understanding how the DGR varies as a function of metallicity is important to have an accurate picture of the gas and dust life cycle in galaxies. How are the heavy metals incorporated into dust and how do metallicity or other local or global parameters control this process? Understanding the behavior of the DGR is also important since many studies determining the dust mass in galaxies then quantify the gas reservoir in galaxies by assuming a DGR. Numerous studies 
 have noted a proportionality of DGR with metallicity, but results can vary depending on the long wavelength data used. \cite[Engelbracht et al (2008)]{engelbracht08} noted a decrease in DGR from the more metal rich galaxies to the moderately metal-poor galaxies, until 12 + log (O/H) $\sim$ 8, beyond which the DGR appears to be constant. Adding 850 \mic\ observations beyond \spit\ wavelengths, however, can increase dust masses (\cite[Galametz et al 2011]{galametz11}). The low metallicity end of the DGR relationship has been particularly ambiguous, since dust mass estimates of dwarf galaxies are sometimes hampered by a submm excess.

For the lowest metallicity galaxies  (12 + log (O/H) $<$ 8.0) where the overall SED peaks at short wavelengths and where we have mostly only upper limits in the submm, we find low upper limits to dust masses and low DGRs compared to the expected value.
The upper limit dust mass from SBS0335-052, for example, is very low (\cite[Galliano et al 2008]{galliano08}; Sauvage et al in preparation) giving an unusually low total DGR, compared to that expected for its extremely low metallicity.
The low DGR measured for the lowest metallicity galaxies may be telling us that perhaps metals are not necessarily incorporated into dust in the same way for the lowest metallicity galaxies as for the more metal rich galaxies. On the other hand, perhaps the total gas reservoir is underestimated.

\begin{figure}[h]
\begin{minipage}[t]{3.5cm}
\vspace{0pt}
\hspace*{5pt}
 \includegraphics[clip,trim=0cm 0.5cm 0cm 0.2cm,width=5.5cm]{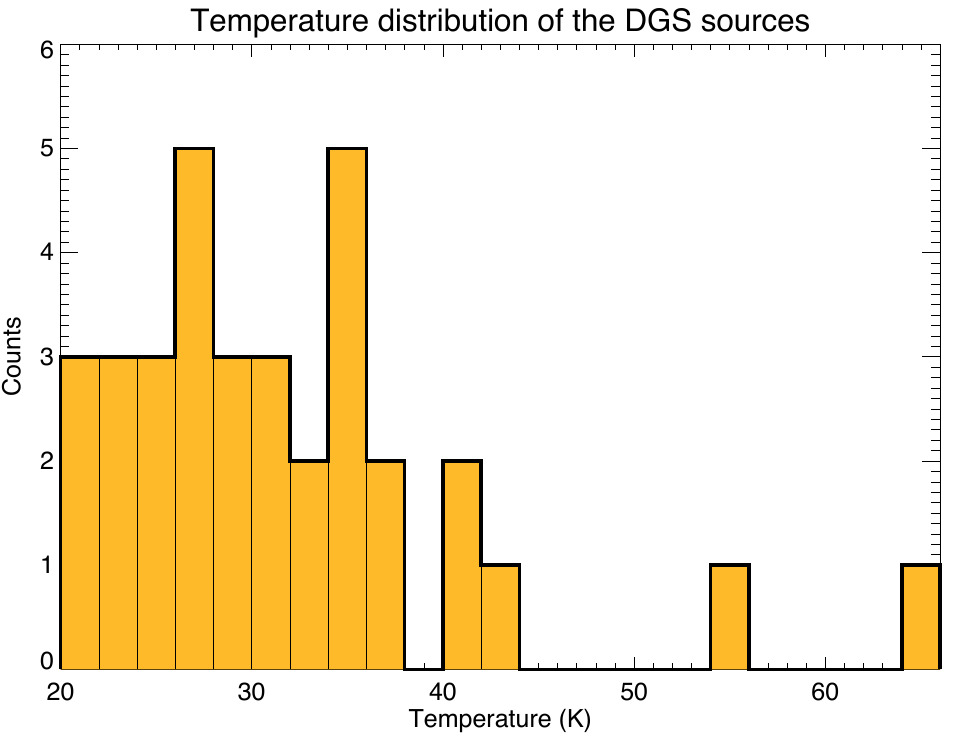}
\hspace*{5pt}
\includegraphics[clip,trim=0cm 0.5cm 0cm 0.2cm,width=5.5cm]{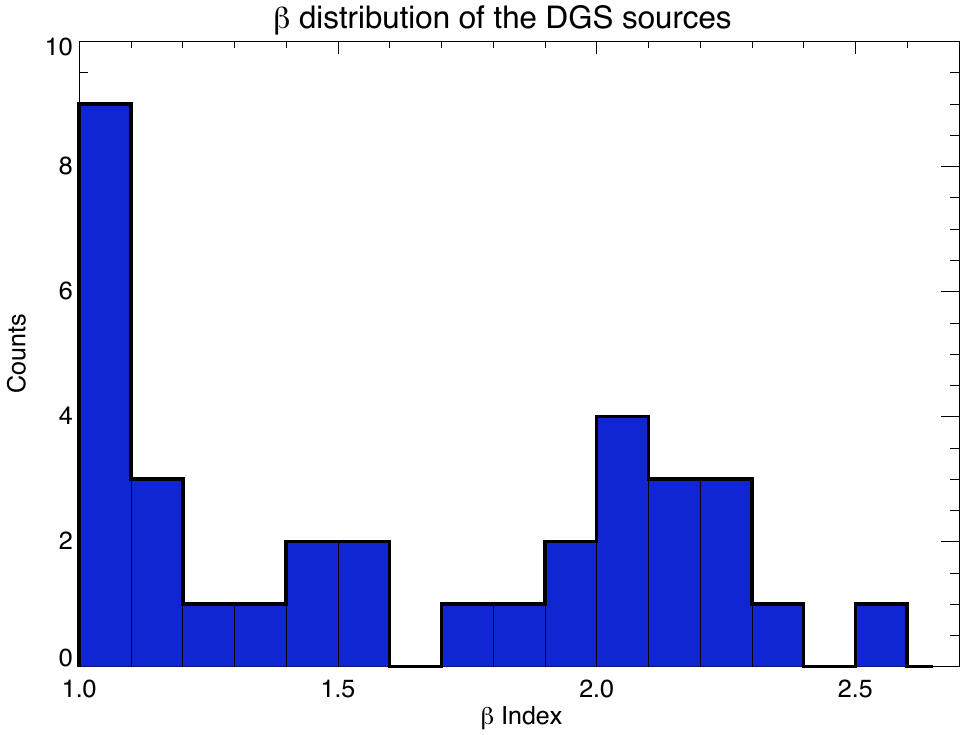}
 \end{minipage}
\
\begin{minipage}[t]{6cm}
\vspace{0pt}
\hspace{80pt}
\includegraphics[width=5.5cm.]{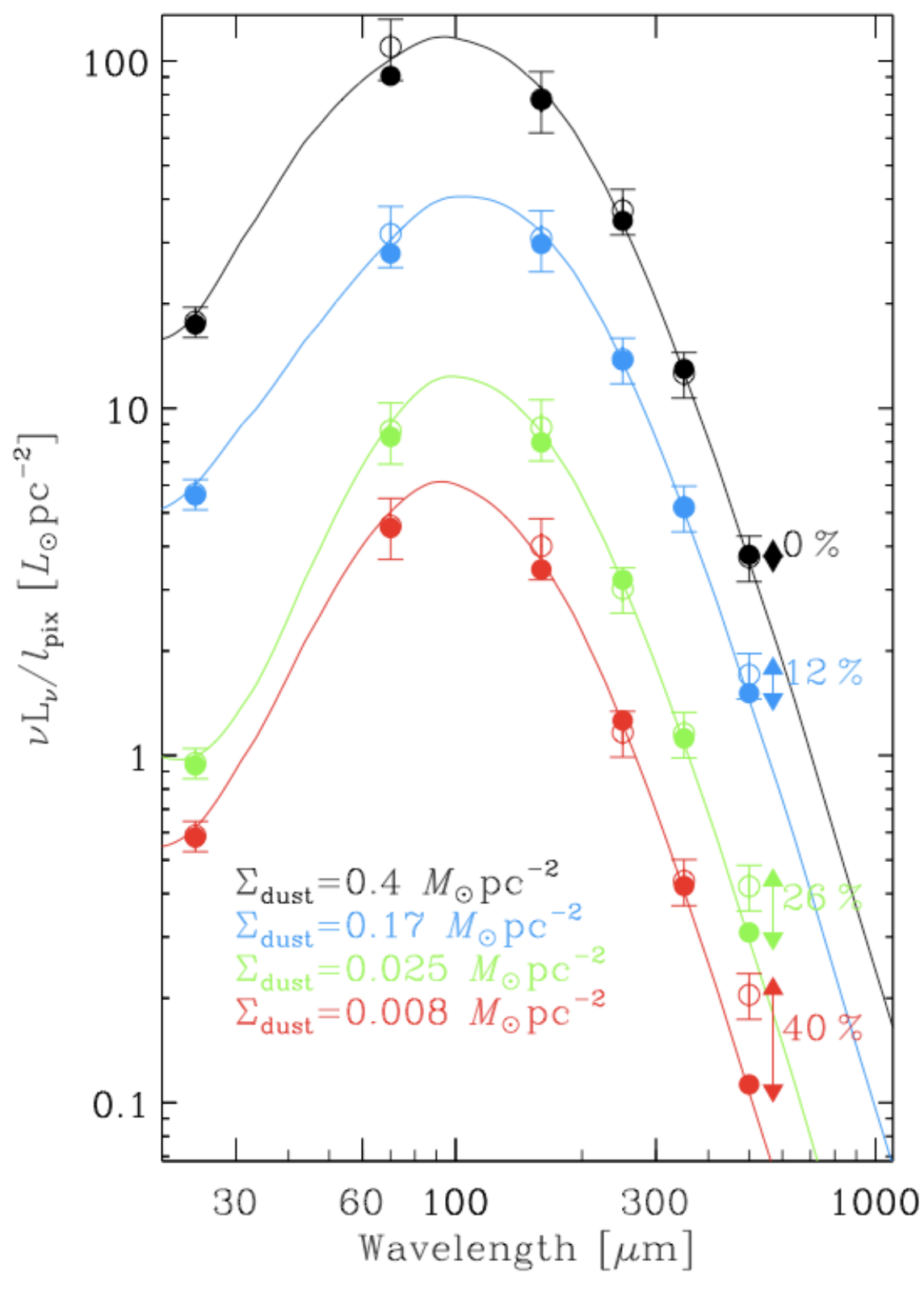} 
\end{minipage}
 \caption{T and  \bet\ distribution of the DGS sample with 500 \mic\ detections.  Selected SEDs within the LMC showing submm excess ranging up to 40\% compared to that expected from the SED models (\cite[Galliano et al 2011]{galliano11}). }
 
\label{fig4}
\end{figure}

While the submm excess has been noted in dwarf galaxies for almost 10 years now - since the first SCUBA observations of dwarf galaxies at 850 \mic\ - the origin remains uncertain.  In the LMC, where Herschel brings 10 pc resolution at 500 \mic, \cite[Galliano et al. (2011)]{galliano11} (this volume) have highlighted locations of submm excess ranging from 15\% to 40\% compared to that expected from the SED models (Fig.\,\ref{fig4}). The excess seems to be anticorrelated with the dust mass surface density. Possible explanations for the submm excess include: 1) very cold dust component (e.g. \cite[Galliano et al 2005]{galliano05}; \cite[Galametz et al 2011]{galametz11};  2) excessive free-free emission; 3) unusual dust emissivity properties (e.g. \cite[Lisenfeld et al 2002]{lisenfeld02}; \cite[M\'eny et al 2002]{meny07}; \cite[Galliano et al 2011]{galliano11}; \cite[Paradis et al 2012]{paradis12}); 4) anomalous spinning dust (e.g. \cite[Draine \& Lazarian 1998]{draine98}; \cite[Ysard et al 2010]{ysard10}).  With more sensitive submm wavelength coverage,  \hers\ is confirming the flater submm slope in more dwarf galaxies. 
Problems exist with some of these explanations put forth for the origin of the excess. For example, invoking a very cold dust component to account for the submm excess can augment the dust mass excessively, resulting in a {\it very high DGR} compared to that expected for the metallicity. But, are we correctly quantifying the total molecular gas mass in low metallicity environments?  
 
\section{The molecular gas in dwarf galaxies}
While CO is the widely used means to access M$_{H_2}$ in galaxies, it is by now a well-established fact that the Galactic X factor, which converts CO to M$_{H_2}$, underestimates the mass of molecular gas in low metallicity galaxies. What is uncertain, however, is how to accurately correct up this factor as a function of metallicity to obtain the total gas reservoir. During the last 2 decades valiant effort has been invested in detecting and interpreting CO emission in low luminosity dwarf galaxies to assess the molecular gas reservoir (e.g. \cite[Leroy et al 2009]{leroy09} and references within).  With the first detections of the 158~\mic\ [CII] line in dwarf galaxies on the KAO, the surprisingly high [CII]/CO values highlighted the fact that CO could be missing a large reservoir of molecular gas (factors of 10 to 100) due to the lower dust abundance. The consequently deeper penetration of UV photons further photodissociate CO, reducing the CO cores - hence, the dearth of detected CO. Due to the self-shielding of H$_{2}$, the larger C$^{+}$-emitting envelope can harbor H$_{2}$  which is not accounted for via CO observations (e.g. \cite[Poglitsch et al 1995]{poglitsch95}; \cite[Madden et al 1997]{madden97}; \cite[Wolfire et al 2010]{wolfire10}). The observed [CII]/CO, thought to be a useful tracer of star formation in galaxies, can sometimes be a factor of a 2 to 5 (or more) times higher in dwarf galaxies than in metal-rich galaxies (e.g. \cite[Madden 2000]{madden00}; \cite[Stacey et al 2010]{stacey10}).

 \begin{figure}
\begin{center}
\includegraphics[width=11 cm.]{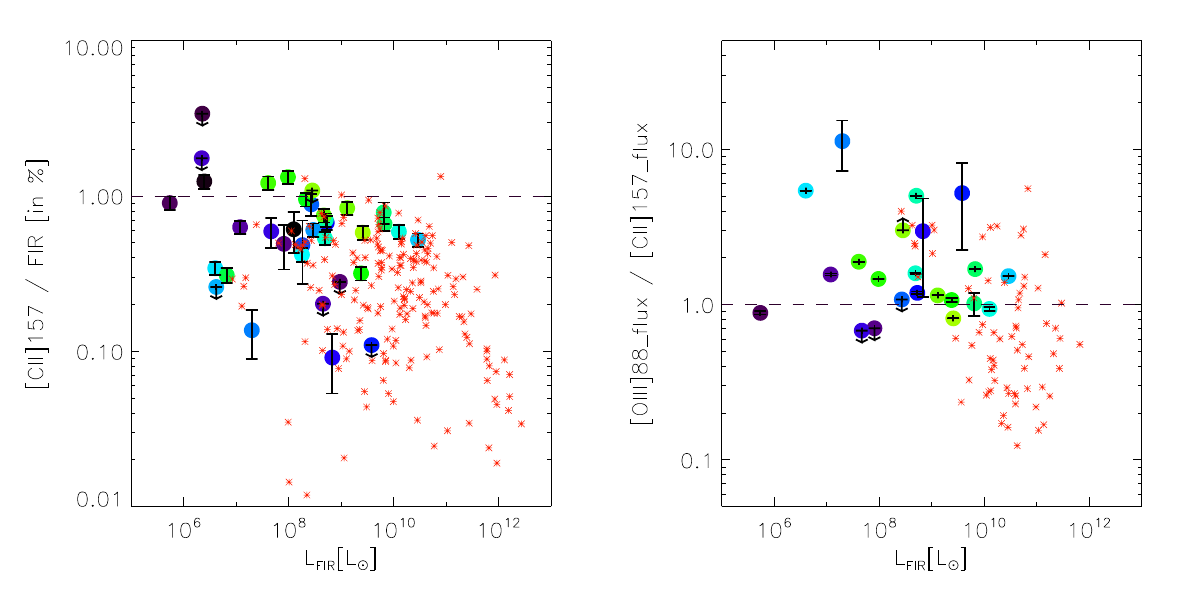} 
\caption{[CII]/FIR and [OIII]/FIR as a function of L$_{[FIR]}$ for the DGS sample (dots) compared to that of \cite[Brauher et al 2008]{brauher08} ISO data, consisting of mostly metal-rich galaxies.} 
  \label{fig5}
\end{center}
\end{figure}

 Dwarf galaxies generally emit a larger fraction of their FIR in the [CII] line, $\sim$ 0.5 to 2 \%, in contrast to the more metal-rich spirals and starbursts which usually show [CII]/FIR less than $\sim$ 0.5\% (Fig.\,\ref{fig5}). If [CII] is the dominant coolant in galaxies and the photoelectric effect is responsible for the primary heating of the ISM, then [CII]/FIR can be thought of as a proxy for the grain photoelectric heating efficiency (e.g. \cite[Rubin et al. 2009]{rubin09}) - the fraction of the power being absorbed by the grains that goes into heating of the ISM. The high [CII]/FIR values observed in dwarf galaxies thus, imply high photoelectric heating efficiencies which may be attributed to properties of the low-metallicity ISM, including lower dust abundance, clumpy ISM, etc. Traversing the galaxy, UV photons suffers less attenuation in the metal-poor ISM, and as a consequence of the longer photon mean-free path,
 the dust on galaxy-wide scales is subject to lower UV flux, effectively decreasing the overall FIR flux and increasing the observed  [CII]/FIR.  The 88 \mic\ [OIII] line is typically the brightest FIR line in dwarf galaxies - usually brighter than the [CII] line (Fig.\,\ref{fig5}), normally considered to be the brightest FIR line in galaxies. This suggests a substantial filling factor of ionised gas in dwarf galaxies and places an important constraint on quantifying the origin of the [CII] line, which also can be excited by electrons in the diffuse ionized gas, not only PDRs (\cite [Lebouteiller et al 2012]{lebouteiller12}). The [CII] and [OIII] together are important calibrators of star formation activity and characterize the diffuse and molecular phases in galaxies. With ALMA, these lines are valuable accessible diagnostics of the ISM of high red-shift galaxies offering new insight on the evolution of the star formation and ISM throughout earlier epochs.

\end{document}